\documentclass[reprint,eqsecnum,floats,aps,amsmath,amssymb,nofootinbib,prd,onecolumn, showpacs]{revtex4-1}

\usepackage{graphicx}
\usepackage{amsmath,amssymb}
\usepackage{hyperref}
\usepackage{graphicx}
\usepackage{subfigure}
\usepackage{arydshln}

\begin{document}

\title{Time-dependent mass of cosmological perturbations in the hybrid and dressed metric approaches to loop quantum cosmology}

\author{Beatriz Elizaga Navascu\'es}
\email{beatriz.b.elizaga@gravity.fau.de}
\affiliation{Institute for Quantum Gravity, Friedrich-Alexander University Erlangen-N{\"u}rnberg, Staudstra{\ss}e 7, 91058 Erlangen, Germany and Instituto de Estructura de la Materia, IEM-CSIC, Serrano 121, 28006 Madrid, Spain}
\author{Daniel Mart\'{\i}n de Blas}
\email{damartind@uc.cl}
\affiliation{Instituto de Estructura de la Materia, IEM-CSIC, Serrano 121, 28006 Madrid, Spain and Instituto de F\'{\i}sica, Facultad de F\'{\i}sica, Pontificia Universidad Cat\'{o}lica de Chile, Av. Vicu\~ {n}a Mackenna 4860, Santiago, Chile}
\author{Guillermo  A. Mena Marug\'an}
\email{mena@iem.cfmac.csic.es}
\affiliation{Instituto de Estructura de la Materia, IEM-CSIC, Serrano 121, 28006 Madrid, Spain}

\begin{abstract}
Loop quantum cosmology has recently been applied in order to extend the analysis of primordial perturbations to the Planck era and discuss the possible effects of quantum geometry on the cosmic microwave background. Two approaches to loop quantum cosmology with admissible ultraviolet behavior leading to predictions that are compatible with observations are the so-called hybrid and dressed metric approaches. In spite of their similarities and relations, we show in this work that the effective equations that they provide for the evolution of the tensor and scalar perturbations are somewhat different. When backreaction is neglected, the discrepancy appears only in the time-dependent mass term of the corresponding field equations. We explain the origin of this difference, arising from the distinct quantization procedures. Besides, given the privileged role that the big bounce plays in loop quantum cosmology, e.g. as a natural instant of time to set initial conditions for the perturbations, we also analyze the positivity of the time-dependent mass when this bounce occurs. We prove that the mass of the tensor perturbations is positive in the hybrid approach when the kinetic contribution to the energy density of the inflaton dominates over its potential, as well as for a considerably large sector of backgrounds around that situation, while this mass is always nonpositive in the dressed metric approach. Similar results are demonstrated for the scalar perturbations in a sector of background solutions that includes the kinetically dominated ones; namely, the mass then is positive for the hybrid approach, whereas it typically becomes negative in the dressed metric case. More precisely, this last statement is strictly valid when the potential is quadratic for values of the inflaton mass that are phenomenologically favored.
\end{abstract}


\maketitle

\section{Introduction}

General relativity (GR) has proven to be a good theoretical framework to describe many phenomena of gravitational origin in the Universe. In particular, combined with quantum field theory (QFT) in curved backgrounds and notions of standard physics, this framework is able to explain, at least to a certain level of accuracy, the evolution of the primordial perturbations that served as seeds for the present large-scale structures and originated the observed cosmic microwave background (CMB) \cite{mukh,langlois}. Despite this success, there is a common belief in the scientific community that, in regimes where the gravitationally relevant quantities approach Planck scale, GR should be corrected. This idea is partially supported by the unavoidable appearance of spacetime singularities in the relativistic theory, such as the big bang \cite{HawElli}. In those regimes, one can argue that a framework which incorporates nonperturbative quantum effects of the geometry (and its interplay with matter) might overcome the breakdown of the classical theory. In the search for this elusive theory of quantum gravity, the identification and analysis of observations that might unveil some traces of those quantum gravitational effects is most important, in order to eventually falsify the theoretical predictions. A possible observational window to such quantum phenomena, or to other alternative modifications of GR, might perhaps be found in the power spectra of the CMB. The increasing precision of the most recent observations \cite{planck,planck-inf} seems to indicate the existence of anomalies in the spectra at large scales. Although the observations at the largest scales are inevitably affected by cosmic variance errors \cite{mukh}, there might exist anomalies even for multipoles of moderately large number, around $\ell \sim 22$ \cite{planck-inf}. This explains the present interest in discussing the potential implications in cosmology of a quantization of the geometry.  

One of the most promising candidates for a theory of quantum gravity is the formalism of loop quantum gravity (LQG). This is a nonperturbative canonical quantization of GR which, in a background-independent way, adopts the strategy followed by Yang-Mills theories, adapted to the description of the gravitational degrees of freedom \cite{lqg}. With an eye on possible tests of the physical consequences that this nonperturbative quantization implies, the techniques of LQG have been applied to cosmological spacetimes, giving rise to a field of research known as loop quantum cosmology (LQC) \cite{lqc1,lqc2,ap,lqc3}. One of the most remarkable results reached for homogeneous spacetimes in LQC is the, quite generic, quantum resolution of the big bang singularity, which becomes what has been named as big bounce \cite{ap,singh}. Specifically, when the (so-called) polymeric quantization that is characteristic of LQG is applied to a homogeneous and isotropic cosmology coupled to a massless scalar field, one obtains families of physical states that are peaked on trajectories that only depart from those of GR when they approach the cosmological singularity, which gets then replaced with a quantum bounce that connects a contracting branch of the Universe with an expanding one \cite{aps,iaps}. Moreover, these trajectories are the solutions of an effective Hamiltonian dynamics that incorporates quantum corrections, leading to a modification of GR that is often called effective LQC \cite{ap,taveras}. The resolution of the initial singularity experimented in LQC opens new avenues to explore the physics of the Early Universe, extending the study of cosmological perturbations beyond the onset of inflation, back to the epochs when the bounce occurred and the eras previous to it, reaching regimes that may be close to the Planck scale. In fact, the effects that the quantum nature of the spacetime may have exerted on the evolution of the cosmological perturbations have been investigated within the context of LQC by a considerable number of authors in recent years, following several different theoretical approaches \cite{hybr-inf1,hybr-inf2,hybr-num, hybr-inf3,hybr-ref,hybr-inf5,hybr-ten,dressed1,dressed2,dressed3,dressed4,effective2,effective3,effective4,effective5}. The possibility that these effects might have left an imprint on the CMB depends on the energy scales associated with the process of inflation and the beginning of the slow-roll regime \cite{Morris,GUI}, details that in turn depend on the type of inflationary spacetimes that are favored in LQC \cite{inflationLQC} (and on some phenomenological parameters). For a recent review of cosmological perturbations in LQC, we refer the reader to Ref. \cite{Edward}.

There exist two approaches to LQC that are based on the combination of a polymeric quantization of the geometry with a Fock quantization of the perturbations: the hybrid approach \cite{hybr-inf1,hybr-inf2,hybr-num,hybr-inf3,hybr-ref,hybr-inf5,hybr-ten} and the dressed metric approach \cite{dressed1,dressed2,dressed3,dressed4}. Their common strategy of combining different types of quantization is based on the assumption that there should exist a regime of physical interest where the main quantum gravity effects come from the homogeneous sector of the cosmology, which is then quantized by means of LQG techniques, while the inhomogeneities can be described with more conventional techniques from QFT in curved backgrounds. This type of strategy was, in fact, first introduced for the quantization of some inhomogeneous Gowdy cosmologies \cite{hybrid1,hybrid2}. The genuine hybrid approach treats the whole system, composed of the background spacetime and its perturbations, as a constrained canonical system, starting from a Hamiltonian formalism that is obtained by truncating the action at quadratic perturbative order \cite{hybr-ref}. The other approach considers the quantization of the homogeneous sector first, it obtains a dressed metric that incorporates the most important quantum corrections within homogeneity, and then lifts the corresponding dynamical trajectories to the truncated phase space that describes the perturbed system at the desired order of approximation \cite{dressed2}.  

In both approaches, the ultraviolet behavior of the perturbations is standard, inasmuch as one gets field equations for the gauge-invariant perturbations that are hyperbolic in the ultraviolet regime \cite{hybr-inf1,hybr-inf2,dressed2}. These field equations for the tensor perturbations and for the so-called Mukhanov-Sasaki invariant \cite{sasa,kodasasa,mukhanov}, which describes the true degrees of freedom of the scalar perturbations, are in fact equivalent to a(n infinite) collection of harmonic oscillators with a time-dependent mass. The quantization of the geometry only alters the value of that time-dependent mass with respect to the standard value in GR. Moreover, the power spectra of the perturbations resulting from these equations have been computed within both approaches, obtaining results that are compatible with the observations \cite{dressed3,GUI}. 

The purpose of this paper is to make clear that, in spite of all the common features shared by the two approaches, they lead in fact to different time-dependent masses, both for the tensor and for the scalar perturbations. As we will see, this is so even when backreaction is neglected and the quantum geometry of the background is described in terms of effective LQC, which is the situation better studied in the literature \cite{dressed3,Morris,GUI}. This difference is important, because it implies that the predictions for the power spectra of the CMB, although similar, are not completely identical. We will explain the reason for this discrepancy, rooted, as expected, in the distinct procedures followed in the two approaches in order to include quantum geometry corrections in the dynamics of the perturbations.
 
In particular, we will study in detail the difference between the time-dependent masses of the two approaches at the bounce. The value and behavior of the mass at the bounce that replaces the singularity in LQC are especially relevant, because this event marks a privileged instant in the evolution of the Universe, and therefore provides a natural choice of time to set some, physically motivated, initial conditions for the perturbations. In fact, these initial conditions are typically understood in cosmology as the definition of an initial vacuum state, both for the tensor and the scalar perturbations. In this sense, the properties of the mass, and more specifically its positivity, can prove very important for the correct definition of those initial conditions. For instance, this would be the case if one wants to construct initial adiabatic vacua for the gauge-invariant perturbations \cite{dressed2,dressed3,dressed4,Morris,GUI,hybr-pred}. 

We will study this positivity, in particular paying an especial attention to background spacetimes in which the energy density of the inflaton at the bounce is dominated by the kinetic contribution. In the two considered approaches these are the most interesting and studied backgrounds \cite{inflationLQC,dressed3,GUI}, because they lead to spectra for the CMB that are compatible with observations and may include quantum geometry corrections. However, we will not restrict the discussion exclusively to those backgrounds. In fact, for spacetimes in which the inflaton energy density is so highly dominated by the kinetic contribution at the bounce that the potential can be safely ignored, the background solutions display a common behavior that can be calculated analytically, as shown in Refs. \cite{waco,wacopre}. This analytic solution can be used to determine the positivity or negativity of the mass at the bounce, provided our considerations are circumscribed to this sector of extreme kinetic dominance.\footnote{We acknowledge an anonymous referee for remarking this point and call attention to the two cited references.} Nonetheless, with the aim at specifying in a quantitative way the regions of solutions where one can assure that the mass turns out to be positive at the bounce, here we want to go beyond that regime of extreme kinetic dominance and complete an analysis that includes sectors of backgrounds with a potential energy density that does not need to be totally negligible. According to the numerical simulations presented in Ref. \cite{waco} this is the case when the parameter that determines the equation of state of the inflaton at the bounce differs from the unit by more than a few percent or, equivalently, when the absolute value of the potential is more than a few percent of the inflaton energy density at the bounce.

The article is structured as follows. In Sec. \ref{sec1}, we first review some basic results about effective homogeneous LQC. Then, we summarize the main ideas that underlie the derivation of the field equations for the gauge-invariant perturbations in the hybrid and dressed metric approaches to LQC, and show the explicit form of the time-dependent masses that appear in those equations when backreaction is neglected and the background obeys the dynamics of effective LQC. At the end of this section, we briefly comment on the difference between the values of these masses in the two considered approaches, identifying the origin of this discrepancy and explaining it on the basis of the different routes to quantization adopted in each case. In Sec. \ref{sec3}, we focus our study on the value of the time-dependent masses at the bounce. We carry out an analytical and numerical study of the properties of these masses, with especial attention paid to the discussion of their positivity. We prove that the mass of the tensor perturbations is always negative (or zero) in the dressed metric approach, and positive in the hybrid approach for a set of background solutions that includes those with kinetically dominated bounces. For scalar perturbations, on the other hand, we demonstrate again the positivity for a sector of backgrounds around the kinetically dominated region in the hybrid case, whereas this is not so generically in the dressed metric approach. In particular, within kinetic dominance and around it, the time-dependent scalar mass of the dressed metric scheme is negative in the relevant case of a quadratic potential when the inflaton mass is not extremely far away from the range of values suggested by phenomenological considerations in LQC \cite{inflationLQC}. Finally, we summarize our results and conclude in Sec. \ref{concl}. Throughout the text, we set the Planck constant $\hbar$ and the speed of light equal to one. Planck units are then defined by taking Newton constant $G$ also equal to one.

\section{Field equations for the gauge-invariant perturbations}\label{sec1}

We start by considering a homogeneous and isotropic, Friedmann-Lema\^{\i}tre-Robertson-Walker (FLRW) spacetime with flat spatial hypersurfaces. For simplicity in the exposition and for mathematical convenience, we will assume compact sections, isomorphic to the three-torus $T^3$, with a compactification length given by a parameter $l_0$. Actually, if this parameter is chosen sufficiently large (much larger than the corresponding Hubble radius), the relevant analyses of cosmological perturbations carried out here become essentially equivalent to those of the noncompact case, which would be reached in a suitable limit with $l_0$ tending to infinity. The FLRW spacetime metric is characterized by a scale factor $a(t)$ and a homogeneous lapse $N_0(t)$. With coordinates adapted to the homogeneity, it can be written as
\begin{align}
ds^{2}=-N_0^2(t)dt^2+a^2(t)\,^0h_{ij} dx^i dx^j,
\end{align}
where $i,j=1,2,3$ denote spatial indices, $^0h_{ij}$ is the Euclidean metric on the compact spatial section, and $x^i$ are periodic Euclidean coordinates, with period equal to $2\pi /l_0$. As a matter source, we minimally couple a homogeneous scalar field $\phi(t)$, with a potential $V(\phi)$, which will play the role of an inflaton field. Indeed, in the classical theory, this scalar field can serve to drive an inflationary period of the geometry. The quantization of this homogeneous and isotropic model, when restricted to a vanishing potential, has been thoroughly studied in LQC \cite{aps,iaps,mmo}. In particular, it is possible to construct a well-defined operator representing the Hamiltonian constraint of the system (see, e.g., \cite{mmo}). Among the solutions to this quantum constraint, it has been shown in a number of analyses \cite{iaps,mop} that there exist states which are highly peaked on trajectories generated by a certain effective Hamiltonian \cite{taveras}, which differs from the classical one by incorporating quantum corrections. Those trajectories have the remarkable property of avoiding the big bang singularity, which gets replaced with a big bounce that connects a contracting branch of the Universe with an expanding one \cite{iaps}. This bounce of quantum origin sets an upper bound on the energy density of the scalar field \cite{ap}.

In the case of interest here of an inflationary cosmology with perturbations, the scalar field that serves as a matter source for the homogeneous sector of the system is not massless, but it is subject to a potential $V(\phi)$. Nonetheless, it is generally admitted (supported in part by the numerical simulations; see, e.g., \cite{ads,lambd}) that the influence of the potential will not change the effective behavior found in the case of the massless field.
The Hamiltonian $H_{|0}^{\rm eff}$ that would generate the effective dynamics of the peaks of these states is then
\begin{align}\label{effh}
N_{0}H_{|0}^{\rm eff}=\frac{N_{0}}{2l_0^3 a^3}\left[\pi_\phi^2- \frac{3l_0^{6}a^{6}}{4\pi G\gamma^2 \Delta} \sin^2 \bigg(\frac{4\pi G\gamma\sqrt{\Delta}\pi_a}{3l_0^3 a^2}\bigg) + 2a^6 l_0^6 V(\phi)\right],
\end{align}
where $\pi_a$ and $\pi_{\phi}$ denote, respectively, the canonically conjugate momenta of $a$ and $\phi$. This effective LQC Hamiltonian, which must vanish on effective solutions, gives rise to the following modified Friedmann and Raychaudhuri equations for the geometry of the inflationary FLRW universe \cite{dressed3,GUI}:
\begin{align}\label{LQC}
\left(\frac{a'}{a}\right)^{2}=\frac{8\pi G}{3}a^2\rho \left(1-\frac{\rho}{\rho_{\rm max}}\right), \qquad \frac{a''}{a}=\frac{4\pi G}{3}a^2\rho\left(1+2\frac{\rho}{\rho_{\rm max}}\right)-4\pi G a^2 P\left(1-2\frac{\rho}{\rho_{\rm max}}\right).
\end{align}
Here, the prime denotes derivative with respect to the conformal time, defined by setting the lapse equal to the effective value of the scale factor, and $\rho$ and $P$ are the energy density and the pressure of the scalar field,
\begin{equation}\label{density}
\rho=\frac{1}{2}\left(\frac{\phi'}{a}\right)^2+V(\phi) ,\qquad P=\rho-2 V(\phi).
\end{equation}
Besides, the quantity $\rho_{\rm max}=3/(8\pi G \gamma^{2}\Delta)$ is the upper bound on the energy density of the inflaton field, where $\gamma$ is the so-called Immirzi parameter \cite{immi} and $\Delta=4\sqrt{3}\pi\gamma G$ is the area gap  allowed by the spectrum of the area operator in LQG \cite{lqg}. When the bound $\rho_{\rm max}$ is reached, the Hubble parameter vanishes and the quantum bounce occurs. Finally, the inflaton field and its momentum satisfy the same relation as in GR,
\begin{align}\label{scalar}
\phi'=\frac{\pi_{\phi}}{l_{0}^3a^{2}}, \qquad \pi_{\phi}'=-l_0^3 a^4 V_{,\phi}.
\end{align}

In order to study small departures from homogeneity, one can now introduce inhomogeneous perturbations of both the metric and the scalar field to lowest nontrivial order in the Einstein-Hilbert action \cite{langlois,HH,shirai,langlois2,pintoneto1,pintoneto2}. According to their properties under symmetry transformations, these perturbations are typically classified as scalar, vector or tensor. In fact, at the considered perturbative order, the vector inhomogeneities are known to be pure gauge when the matter content is a scalar field. Therefore, we will devote most of our discussion to the scalar and tensor perturbations. It is convenient to expand these perturbations in a complete set of scalar, vector, and tensor harmonics. Specifically, given the Euclidean metric $^0h_{ij}$ and its associated affine connection, one can consider a complete set of eigenfunctions of its Laplacian, which can be understood as plane waves with wave vectors $\vec{k}$ that, in the compact case, are integer tuples multiplied by $2\pi/l_0$. From these eigenfunctions, as well as forming appropriate combinations of the metric $^0h_{ij}$ with its associated connection, one can construct the desired complete set of scalar, vector, and tensor harmonics \cite{Bardeen}. Let us notice at this point that, thanks to the compactness of the spatial sections, in all these harmonic expansions we can and will exclude their zero-modes, and regard them as part of the homogeneous metric and matter variables.

\subsection{Hybrid quantization approach}

Let us begin by summarizing the main ideas that underlie the hybrid (LQC) approach to the quantization of cosmological perturbations. For specific details about the derivation of the equations discussed here, we refer the reader to Refs. \cite{hybr-inf1,hybr-inf2,hybr-inf3,hybr-ref,hybr-inf5,hybr-ten}. The strategy followed in those works is, in short, to truncate the Einstein-Hilbert action at quadratic order in the perturbations, and then regard the entire truncated system as a constrained symplectic manifold. The Hamiltonian that results from this truncation is a linear combination of constraints, which capture the covariance of the relativistic system up to the order of the truncation. More specifically, this Hamiltonian is the sum of the following terms. On the one hand, the homogeneous lapse function $N_0$ is the Lagrange multiplier of the zero-mode of the Hamiltonian constraint, which is in turn formed by the Hamiltonian constraint of an FLRW model plus an infinite sum of functions that are quadratic in the mode coefficients of the scalar and tensor perturbations. On the other hand, the infinite number of mode coefficients of the perturbations of the lapse and the shift vector serve as the Lagrange multipliers of the linearization of the Hamiltonian constraint and of the momentum constraints of GR, respectively.

This truncated cosmological system can be recast in terms of gauge-invariant canonical variables for the perturbations, as shown in Ref. \cite{hybr-ref}. The advantages of a description in terms of quantities that are invariant under the transformations generated by the linear perturbative constraints is evident. The main steps of this reformulation are as follows. First, with the considered matter content, the tensor perturbations of the metric are automatically gauge-invariant.  On the other hand, by appropriately combining the scalar perturbations of the metric with those of the inflaton field, one can obtain the Mukhanov-Sasaki gauge-invariant field \cite{sasa,kodasasa,mukhanov}. Both of these tensor and scalar invariants still are so if one rescales them with a function of the homogeneous scale factor (and their conjugate momenta with one over this function, up to the sum of terms that depend only on the configuration variables). Among the variables allowed by this freedom in the choice of gauge invariants, the hybrid approach fixes those which have configuration mode coefficients that, classically, obey the second-order equation of a harmonic oscillator with a time-dependent mass and without any friction term \cite{mukh,hybr-ref,hybr-ten}. We will call $\tilde{d}_{\vec k,\epsilon}$ the corresponding configuration mode coefficients for these tensor gauge invariants, where $\epsilon$ is a dichotomy label that denotes each of the two possible polarizations. On the other hand, let $v_{\vec k}$ be the configuration mode coefficients of the chosen Mukhanov-Sasaki gauge invariant. Apart from the advantages of describing the primordial fluctuations with these specific variables in standard cosmology \cite{mukh} (given the almost-Gaussian distribution of the anisotropies in the CMB), they turn out to be the only ones, among those related by the mentioned rescaling transformations, for which the dynamics can be unitarily implemented when one adopts an adequate Fock quantization for them \cite{unique1,unique2,unique3,unique3b,fmov,uniqueds,uniquesignature,unique4}, in the context of QFT in curved spacetimes. This feature is appealing in view of the hybrid ideas for the later quantization of the system, which employs a Fock representation of the perturbations. Once these gauge-invariant canonical variables have been chosen, the rest of information contained in the perturbative sector of the phase space can be codified in the, conveniently Abelianized, linear perturbative constraints, together with their canonical momenta \cite{hybr-ref}. Now, let us recall that the hybrid scheme demands a canonical formulation of the whole system formed by the homogeneous degrees of freedom and the perturbations. Therefore, since the chosen tensor and Mukhanov-Sasaki gauge invariants involve, in their definition in terms of the metric and inflaton perturbations, the homogeneous canonical variables, these homogeneous quantities must be corrected by terms quadratic in perturbations so that the new, corrected, homogeneous variables complete the gauge invariants and the linear perturbative constraints, together with their momenta, into a canonical set for the entire cosmological system \cite{hybr-ref,hybr-ten}. The Hamiltonian for the whole cosmology is then expressed in terms of this new set of canonical variables, prior to its quantum representation.

To implement the quantization, the hybrid approach combines some quantum gravity inspired representation (in this paper it will be LQC) of the homogeneous sector of the phase space with a more conventional Fock representation of the rest of inhomogeneous degrees of freedom \cite{hybrid1,hybrid2,hybrid3}. The kinematic representation space of the quantum theory is the tensor product of the different Hilbert or Fock spaces associated with these sectors. One then constructs a quantum representation of the different constraints that the relativistic system possesses and imposes them following the Dirac procedure \cite{Dirac}. In the studied perturbed cosmological spacetimes, since the linear perturbative constraints are part of the selected canonical variables, their quantum imposition is straightforward: they just restrict the physical states not to depend on their conjugate momenta. On the other hand, the quantum imposition of the zero-mode of the Hamiltonian constraint is a highly nontrivial task. In particular, the perturbative contributions to this constraint couple the homogeneous sector with the Mukhanov-Sasaki and tensor perturbations \cite{hybr-ref}.

In order to find solutions of cosmological interest to this complicated quantum constraint, the following ansatz has been proposed \cite{hybr-inf3,hybr-ref,hybr-ten}. One considers wave functions in which the dependence on the different sectors of the phase space factorizes, except for the homogeneous inflaton, which then may be viewed as an internal time for these quantum states. In particular, the homogeneous part of these states, $\Gamma (a,\phi)$, where $a$ generically denotes dependence on the homogeneous geometry, may be chosen as an exact solution to the, homogeneous, quantum FLRW model, and in this paper we will take it this way. However, let us comment that this choice is in principle not needed, and in fact one may consider other possibilities for $\Gamma$ that incorporate the presence of some quantum backreaction of the perturbations onto the homogeneous sector of the model \cite{hybr-ref,hybr-ferm}. With this ansatz at hand, and provided that $\Gamma$ is sufficiently peaked on the homogeneous geometry for all values of $\phi$, then the imposition of the zero-mode of the Hamiltonian constraint leads to the requirement that the outcome of certain operators, acting exclusively on the Mukhanov-Sasaki and the tensor parts of the wave function, must be zero. Remarkably, these quantum equations on the gauge-invariant perturbations only depend on the homogeneous geometry via some expectation values of geometric operators taken on $\Gamma$ \cite{hybr-ref,hybr-ten}. On the other hand, as expected, the quantum dependence on the Mukhanov-Sasaki and tensor configuration variables, as well as on their momenta, is quadratic. Therefore, it seems reasonable that, for some of our considered states, one can legitimately substitute this quadratic dependence on the Mukhanov-Sasaki and tensor variables by its classical counterpart, and then regard the operators that act on the perturbative parts of the wave function as constraints that incorporate the effect of quantum geometry contributions. In this situation, one can easily obtain the dynamical equations for the Mukhanov-Sasaki and tensor perturbations with quantum geometry corrections \cite{hybr-ref,hybr-ten}. Furthermore, in the considered scenario with negligible backreaction, we may choose $\Gamma$ to be highly peaked on the effective LQC trajectories generated by $H_{|0}^{\rm eff}$, given in Eq. \eqref{effh}. If this is the case, we may substitute that effective behavior in the expectation values of the geometry which appear in the previous dynamical equations, arriving at the following evolution for the tensor and the Mukhanov-Sasaki perturbations \cite{hybr-pred,GUI},
\begin{align}\label{h-efft2}
\tilde d_{\vec {k},\epsilon}''+\left[k^2-\frac{4\pi G}{3}a^2(\rho-3P)\right]\tilde d_{\vec {k},\epsilon}=0,\\\label{h-effs2}
v_{\vec k}''+\left[k^2-\frac{4\pi G}{3}a^2(\rho-3P)+\mathcal{U}\right]v_{\vec k}=0,
\end{align}
where the prime denotes again derivative with respect to the conformal time, and
\begin{align}\label{Ueff}
\mathcal{U}=a^2\left[V_{,\phi\phi}+48\pi G V(\phi)+ 6\frac{a'{\phi}'}{ a^{3}\rho}V_{,\phi}-\frac{48\pi G}{\rho} V^2(\phi)\right].
\end{align}

In order to compute this Mukhanov-Sasaki potential $\mathcal{U}$, one needs to provide certain prescriptions for the quantum representation of the functions of the homogeneous variables that couple to the perturbations in the full Hamiltonian constraint. For the effective LQC approximation considered here, the only relevant one of all such prescriptions consists of adjusting the length of the holonomies which encode the information about the Hubble parameter in LQC. This adjustment is made to preserve the superselection sectors in which the Hamiltonian constraint of the homogeneous model separates the kinematic Hilbert space \cite{hybr-inf3}. 

Finally, it is worth noticing that, in those regimes where the effective dynamics of the geometric background approaches classical linearized GR [so that one can ignore quadratic terms in the pressure and density in Eq. \eqref{LQC}], the $k$-independent term that appears in both the tensor and the Mukhanov-Sasaki equations approaches the classical quantity $-a''/a$, thus recovering the well-known classical equations for the gauge-invariant perturbations.

\subsection{Dressed metric approach}

Let us now summarize the main features of the dressed metric approach to the quantization of cosmological perturbations within the context of LQC. For specific details about the arguments and derivations of the corresponding equations, we refer the reader to Refs. \cite{dressed1,dressed2,dressed3,dressed4}. First of all, let us point out that this approach follows the same hybrid quantization strategy of combining LQC techniques for the homogeneous degrees of freedom with Fock representations for the perturbations. However, there is a major difference between the two approaches: in the dressed metric case, one does not regard the truncated, perturbed cosmological system as a full symplectic and constrained manifold. One treats in a separate way the homogeneous background and the inhomogeneous perturbations, assuming since the very beginning that backreaction effects should be ignorable. In particular, one deals with the phase space evolution in two steps: one first obtains the dynamical trajectories on the homogeneous sector and then lifts them to the truncated phase space \cite{dressed2}. Consequently, in this approach, one lacks a classical Hamiltonian that generates the evolution of both the homogeneous background and the perturbations, at the considered order of truncation. Instead, one may understand the dressed metric formalism as if it possessed two different Hamiltonians. The first one is just the standard FLRW Hamiltonian. The second one is the Hamiltonian that, when the homogeneous background is viewed as a fixed entity, gives rise to the linearized equations for the perturbations \cite{dressed2,dressed3}. 

The perturbative sector of the cosmological system is again given in terms of gauge-invariant quantities, although their description is somewhat different from the one put forward in the hybrid approach. In the dressed metric approach, one solves classically the linear perturbative constraints. The resulting, reduced, phase space for the perturbations is then described with a specific choice of tensor and Mukhanov-Sasaki variables. For the tensor degrees of freedom, we will follow the notation of Ref. \cite{dressed2} and call  $T^{(\epsilon)}_{\vec k}/l_{0}^{3}$ the configuration mode coefficients of the tensor perturbations (where $\epsilon$ denotes again the polarization). In turn, $\mathcal{Q}_{\vec k}/l_{0}^{3}$ will denote the configuration mode coefficients of the Mukhanov-Sasaki field variable chosen in the dressed metric approach.

As we have commented, the philosophy to quantize the system is to combine an LQC representation for the homogeneous sector of the (truncated) phase space with a Fock representation for the tensor and Mukhanov-Sasaki perturbations, as in the hybrid approach. Again, one also introduces an ansatz for the quantum states in which the dependence on the homogeneous geometry and on the perturbations factorizes. In this ansatz, all partial wave functions are allowed to depend on the inflaton $\phi$, which is viewed as an internal time. However, in the dressed  metric case there is no Hamiltonian constraint that affects the perturbations, since the whole of the truncated cosmology is not treated as a constrained symplectic system. Instead, one has the Hamiltonian constraint of the homogeneous FLRW model, and the Hamiltonian functions that, classically, generate the dynamics of the perturbations. Accordingly, the approach requires the homogeneous part of the states to be an exact solution of the FLRW model in LQC, and then uses this solution to define the quantum dynamics on the phase space of the gauge-invariant perturbations \cite{dressed2,dressed3}. In this way, the perturbations behave as test fields that see a dressed metric determined by certain expectation values of operators of the homogeneous geometry, which incorporate the most relevant quantum effects. In our case with compact sections, one can construct in this manner, for instance, operators representing the Hamiltonians on the phase space of the perturbations. Associated with these operators, one obtains Schr\"odinger equations in $\phi$ for the partial wave functions that describe the tensor and the Mukhanov-Sasaki perturbations. 

One may again consider in this approach that the homogeneous part of the wave function is highly peaked on a trajectory dictated by the effective LQC dynamics generated by $H_{|0}^{\rm eff}$. Such an effective description would then also apply to the dressed metric quantities that couple to the scalar and tensor perturbations. The field equations of the tensor and Mukhanov-Sasaki variables propagating on such effective dressed metric are \cite{dressed2,dressed3}
\begin{align}
T_{\vec k}^{(\epsilon)\prime \prime}+2\frac{a'}{a}T_{\vec k}^{(\epsilon)\prime}+k^{2}T_{\vec k}^{(\epsilon)}=0,\\
\mathcal{Q}_{\vec k}''+2\frac{a'}{a}\mathcal{Q}_{\vec k}'+(k^{2}+\mathcal{V})\mathcal{Q}_{\vec k}=0,
\end{align}
with the same notation as before for the prime and where all the homogeneous quantities must be evaluated on the effective LQC background [see Eq. \eqref{LQC}]. Besides,
\begin{align}\label{V}
\mathcal{V}=\left[\mathfrak{f}V(\phi)-2\sqrt{\mathfrak{f}} V_{,\phi}+ V_{,\phi\phi}\right]a^2, \qquad \mathfrak{f}=\frac{48\pi G \pi_{\phi}^{2}}{\pi_{\phi}^2 +l_{0}^{6}a^{6}V(\phi)},
\end{align}
a function that can be checked to coincide with the hybrid Mukhanov-Sasaki potential $\mathcal{U}$ in an FLRW universe, and therefore at the order of truncation adopted in the dressed metric approach. A caveat is in order here: the coefficient of $V_{,\phi}$ in $\mathcal{V}$, when evaluated on classical FLRW solutions, equals $-12a|\pi_{\phi}|/|\pi_{a}|$ if the square root of $\mathfrak{f}$ is defined as positive, or $12a|\pi_{\phi}|/|\pi_{a}|$ if it is defined as negative. On the other hand, the corresponding coefficient in $\mathcal{U}$, which is in fact the one that appears in the context of linearized GR \cite{mukh}, is given by $-12a\pi_{\phi}/\pi_{a}$. This tension may be solved by demanding that the sign of the square root of $\mathfrak{f}$ be positive when the signs of $\pi_{\phi}$ and $\pi_{a}$ coincide, and negative otherwise. Alternatively, in recent analyses on the consequences of the dressed metric approach for the CMB \cite{waco,waco1}, the considered coefficient of $V_{,\phi}$ in the potential $\mathcal{V}$ has been taken equal to $2a^{2}\sqrt{24\pi G}\dot{\phi}\rho^{-1/2}$, where the dot denotes derivative with respect to the proper time. One can see that, classically, this last expression coincides for an expanding universe with the result obtained in linearized GR, and thus also with the hybrid result.

Now, if one compares the Hamiltonian functions for the perturbations in the dressed metric approach with the ones that generate the evolution of these perturbations in the hybrid approach \cite{hybr-ref,hybr-ten,dressed1,dressed2}, one can see that the choices of variables employed in each of these approaches for the description of the tensor and Mukhanov-Sasaki perturbations can be related by a very specific transformation, which is canonical as far as the perturbations are concerned. In particular, this transformation involves the multiplication of the configuration variables $T_{\vec k}^{(\epsilon)}$ and $\mathcal{Q}_{\vec k}$ by the homogeneous scale factor of the cosmology (up to a constant). So, in order to compare the dressed field equations for the perturbations with the ones obtained in the hybrid quantization approach for the tensor and Mukhanov-Sasaki variables $\tilde d_{\vec {k},\epsilon}$ and $v_{\vec k}$, it is most convenient to consider their analogues for
\begin{align}
t_{\vec k}^{(\epsilon)}=\frac{a}{\sqrt{32\pi Gl_{0}^{3}}}T_{\vec k}^{(\epsilon)} \qquad \text{and} \qquad q_{\vec k}=\frac{a} {\sqrt{l_{0}^{3}}}\mathcal{Q}_{\vec k},
\end{align}
where $a$ corresponds again to the effective, dressed scale factor. Their equations, after evaluating the resulting explicit time derivative $a^{\prime\prime}$ on effective LQC trajectories [via Eq. \eqref{LQC}], are
\begin{align}\label{d-efft2}
t_{\vec{k}}^{(\epsilon)\prime\prime}+\left[k^{2}-\frac{4\pi G}{3} a^2 \rho\left(1+2\frac{\rho}{\rho_{\rm max}}\right)+4\pi G a^2 P\left(1-2\frac{\rho}{\rho_{\rm max}}\right)\right]t_{\vec k}^{(\epsilon)}=0,\\\label{d-effs2}
q_{\vec{k}}''+\left[k^{2}-\frac{4\pi G}{3} a^2 \rho\left(1+2\frac{\rho}{\rho_{\rm max}}\right)+4\pi G a^2 P\left(1-2\frac{\rho}{\rho_{\rm max}}\right)+\mathcal{V}\right]q_{\vec k}=0,
\end{align}
with
\begin{align}\label{Veff}
{\mathcal{V}}=a^2\left[V_{,\phi\phi}+48\pi G V(\phi)-\text{sign}\left(\sqrt{\mathfrak{f}}\right)\frac{4\sqrt{6\pi G}|\phi'|}{a\rho^{1/2}}V_{,\phi}-\frac{48\pi G}{\rho} V^2(\phi)\right].
\end{align}

In those regimes in which the effective dynamics of the background approaches that of linearized GR (and when the sign of the square root of $\mathfrak{f}$ is appropriately taken in $\mathcal{V}$, as we have commented above), these equations for the perturbations coincide with the hybrid ones and, in turn, with the classical tensor and Mukhanov-Sasaki equations.

\subsection{Differences in the perturbation dynamics within effective LQC}

Owing to the different quantization strategies adopted in the hybrid and dressed metric approaches, as we have explained above, the following differences arise in the perturbations equations derived from them in effective LQC:
\begin{itemize}
\item The $k$-independent term that appears in both the tensor and the Mukhanov-Sasaki equations, and which equals $-a''/a$ in classical GR, is not the same in the two approaches [see Eqs. \eqref{h-efft2}, \eqref{h-effs2}, \eqref{d-efft2}, and \eqref{d-effs2}]. This can be traced back to the differences in the treatment of the phase space of the perturbed FLRW cosmologies in the hybrid and the dressed metric approaches. In the hybrid case, the whole phase space is treated as a symplectic manifold. The $k$-independent factor is then expressed in terms of canonical variables. It is the expectation value of the operator representing this canonical expression what is evaluated on trajectories described by the effective dynamics of LQC. On the contrary, in the dressed metric formalism, one does not have a global canonical symplectic structure on the truncated phase space. The dressed term $-a''/a$ is evaluated on effective solutions to LQC, including the computation of the time derivatives, which are calculated along the trajectories of the effective dynamics. The difference then arises because of the departure between the standard classical relation of the time derivatives of the scale factor with its canonical momentum and the corresponding effective relation in LQC. In other words, given that $a''/a=\{a\{a,H_{|0}\},H_{|0}\}$ classically, where $H_{|0}$ is the Hamiltonian constraint of the inflationary FLRW cosmology, the difference appears because
\begin{align}
\left(\{a\{a,H_{|0}\},H_{|0}\}\right)_{\rm eff}\neq \{a\{a,H^{\rm eff}_{|0}\},H^{\rm eff}_{|0}\},
\end{align}
where the subscript ``eff'' indicates evaluation on effective solutions after having computed the Poisson brackets.

\item The Mukhanov-Sasaki potentials $\mathcal{U}$ and $\mathcal{V}$ are different as well. Leaving aside a subtlety concerning the sign of the contribution of $V_{,\phi}$ in $\mathcal{V}$, related with the way in which $\sqrt{\mathfrak{f}}$ is chosen and which is present even if the effective dynamics reproduces classical GR, the two potentials indeed display discrepancies in effective LQC. Specifically, if we compare the expression \eqref{Veff} of $\mathcal{V}$ with the hybrid potential $\mathcal{U}$ given in Eq. \eqref{Ueff}, we observe that they differ in the absolute value of the factor that multiplies $V_{,\phi}$, where it is especially remarkable the absence of the first time derivative of the (effective) scale factor in the dressed metric case. The discrepancy arises, once more, owing to the different quantization prescriptions followed in the hybrid and the dressed metric approaches. In particular, in the hybrid approach, one is naturally led to adopt a specific prescription that preserves the superselection sectors of the homogeneous geometry, since the potential $\mathcal{U}$ is part of a constraint operator acting on the entire quantum space that describes both the homogeneous cosmology and the perturbations. This is not the case in the dressed metric approach, where one just evaluates $\mathfrak{f}$ [given in Eq. \eqref{V}] on effective LQC trajectories and then takes its square root (as it is directly proposed in Refs. \cite{dressed2,dressed3}).
\end{itemize}

It is worth noting, nonetheless, that the latter difference between the two Mukhanov-Sasaki potentials $\mathcal{U}$ and $\mathcal{V}$ is only expected to be relevant in regimes where the energy density of the scalar field is not kinetically dominated, since it is only then that the effect of the potential, and hence the contribution of $V_{,\phi}$, can be important. But a kinetic dominance would precisely be the case at the bounce for the most interesting effective solutions, since a physically acceptable period of slow-roll inflation compatible with the persistence of quantum geometry effects on the largest scales observed in the CMB typically requires solutions of this type \cite{GUI}. However, since in the passage from the bounce to the inflationary regime, the contribution of the potential becomes significant on those effective solutions, the commented difference might not be totally ignorable during some stages of the evolution. 

\begin{figure}
	\includegraphics[width=0.49\textwidth]{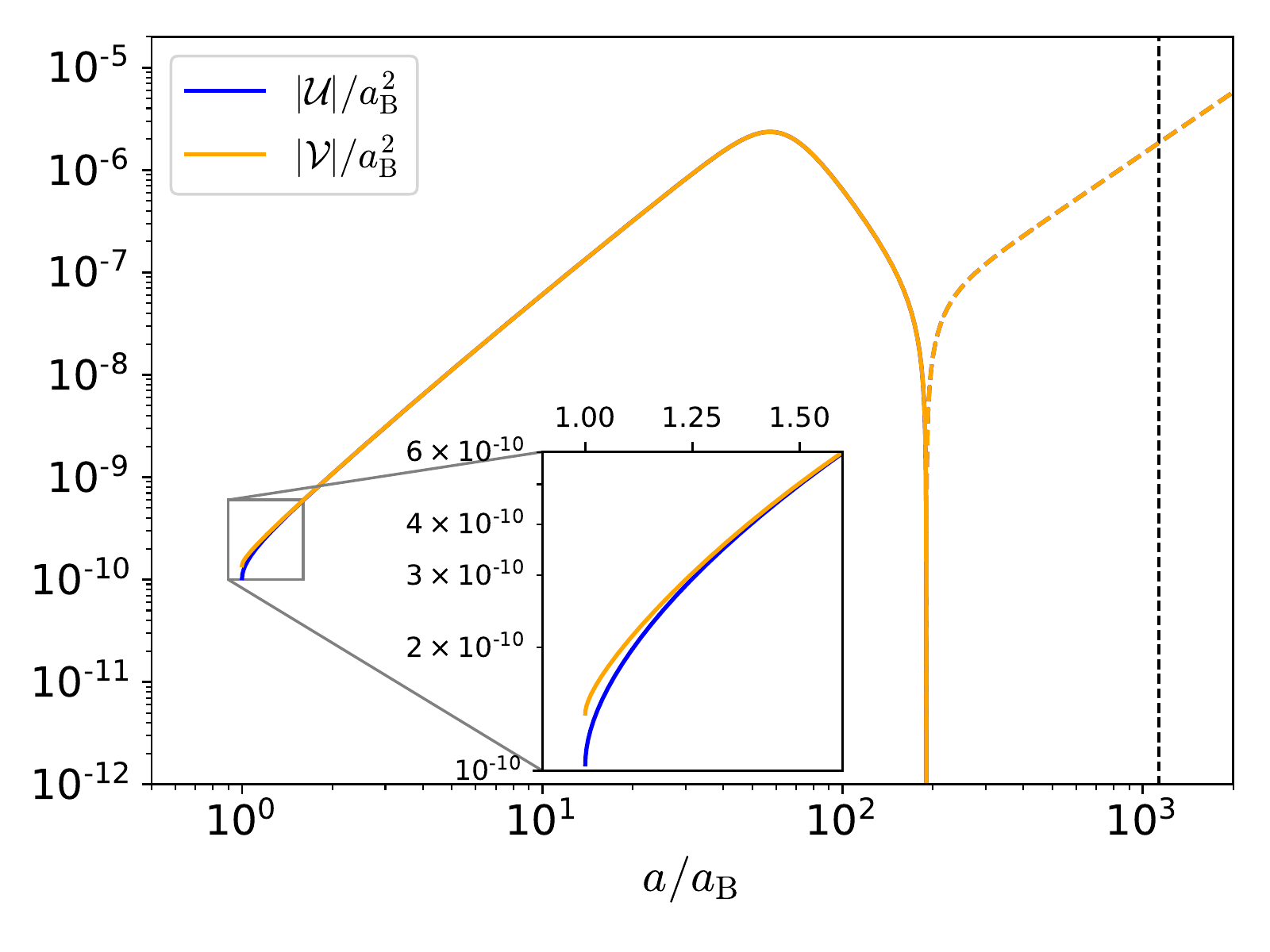}
	\includegraphics[width=0.49\textwidth]{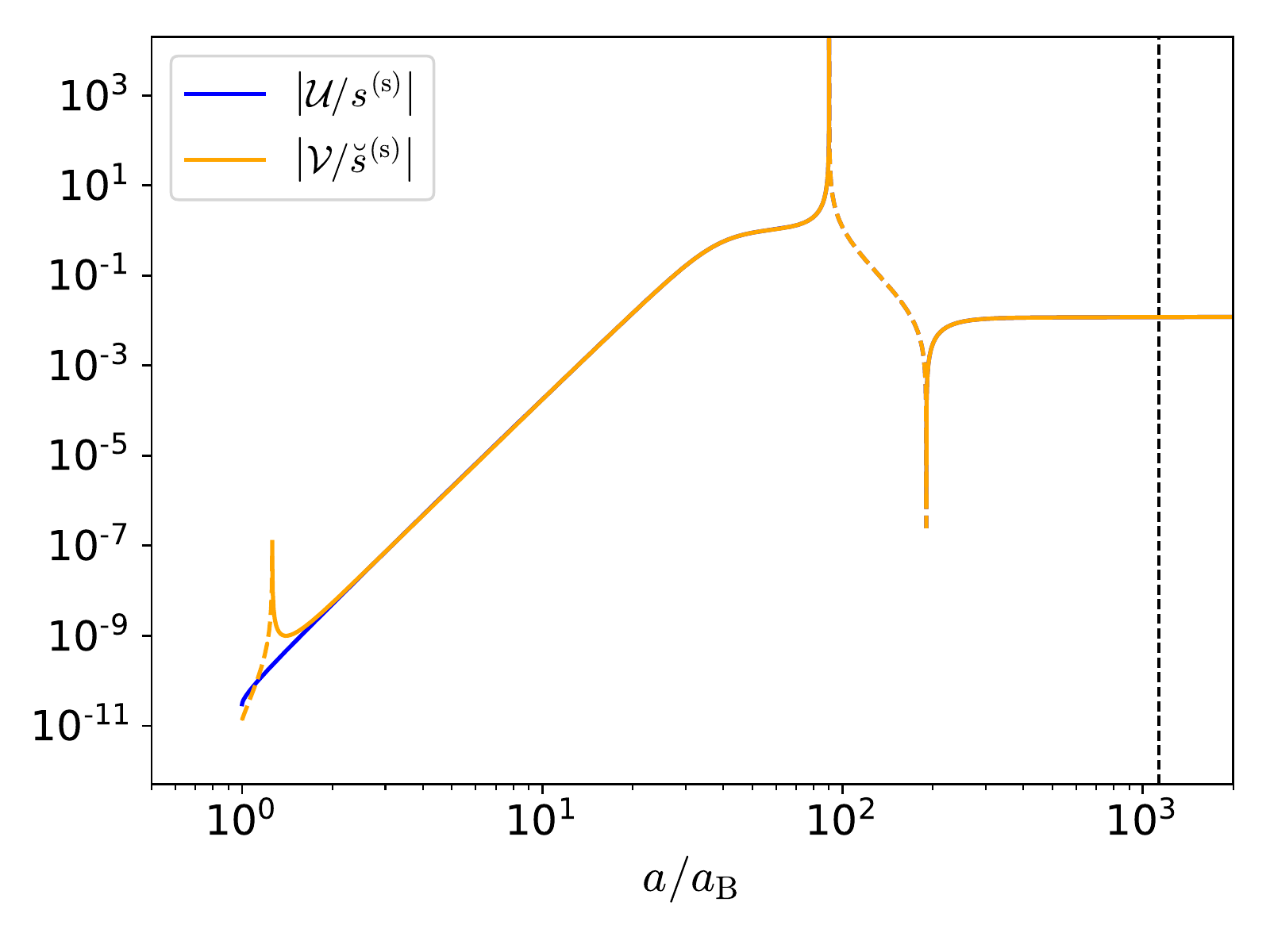}
	\caption{Left panel: Evolution, on the expanding branch of the Universe, of the Mukhanov-Sasaki potentials $\mathcal{U}$ and $\mathcal{V}$, corresponding respectively to the hybrid and dressed metric approaches. Right panel: Relative contribution of the Mukhanov-Sasaki potential to the corresponding time-dependent mass, denoted by $s^{({\rm s})}$ in the hybrid approach and by $\breve{s}^{({\rm s})}$ in the dressed metric approach. Here, we consider a quadratic potential, $V(\phi) = m^{2}\phi^2/2$. In Planck units, the inflaton field at the bounce and the parameters of the model are taken equal to $\phi_{\rm B} = 0.97$, $m =1.20\cdot 10^{-6}$, and $\gamma = 0.2375$. We represent the absolute value of the considered quantities, distinguishing between positive and negative values by employing solid and dashed lines, respectively. The black vertical dashed line marks the onset of the slow-roll phase.}
	\label{fig:uvpot}
\end{figure}

\begin{figure}
	\includegraphics[width=0.49\textwidth]{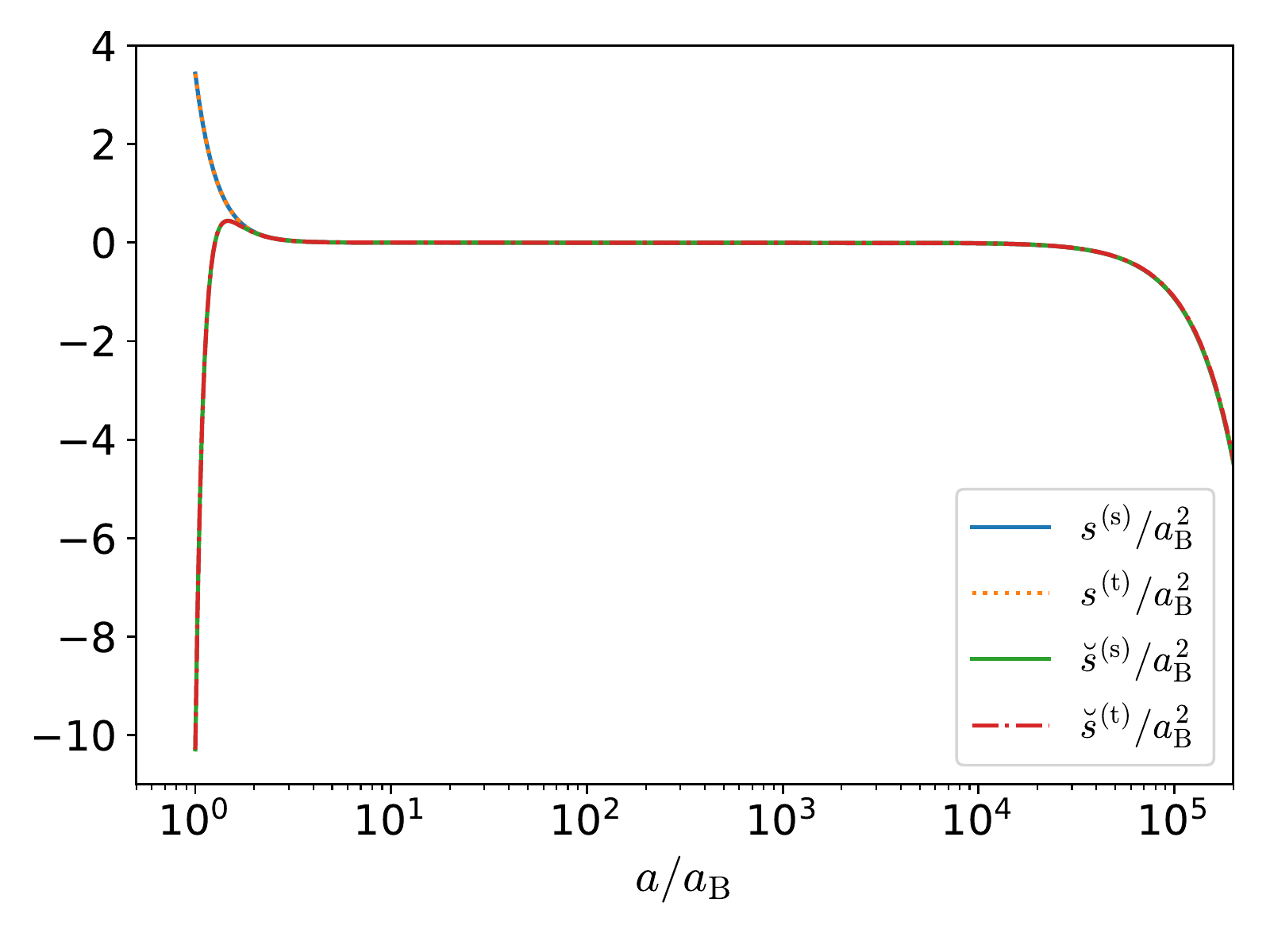}
	\includegraphics[width=0.49\textwidth]{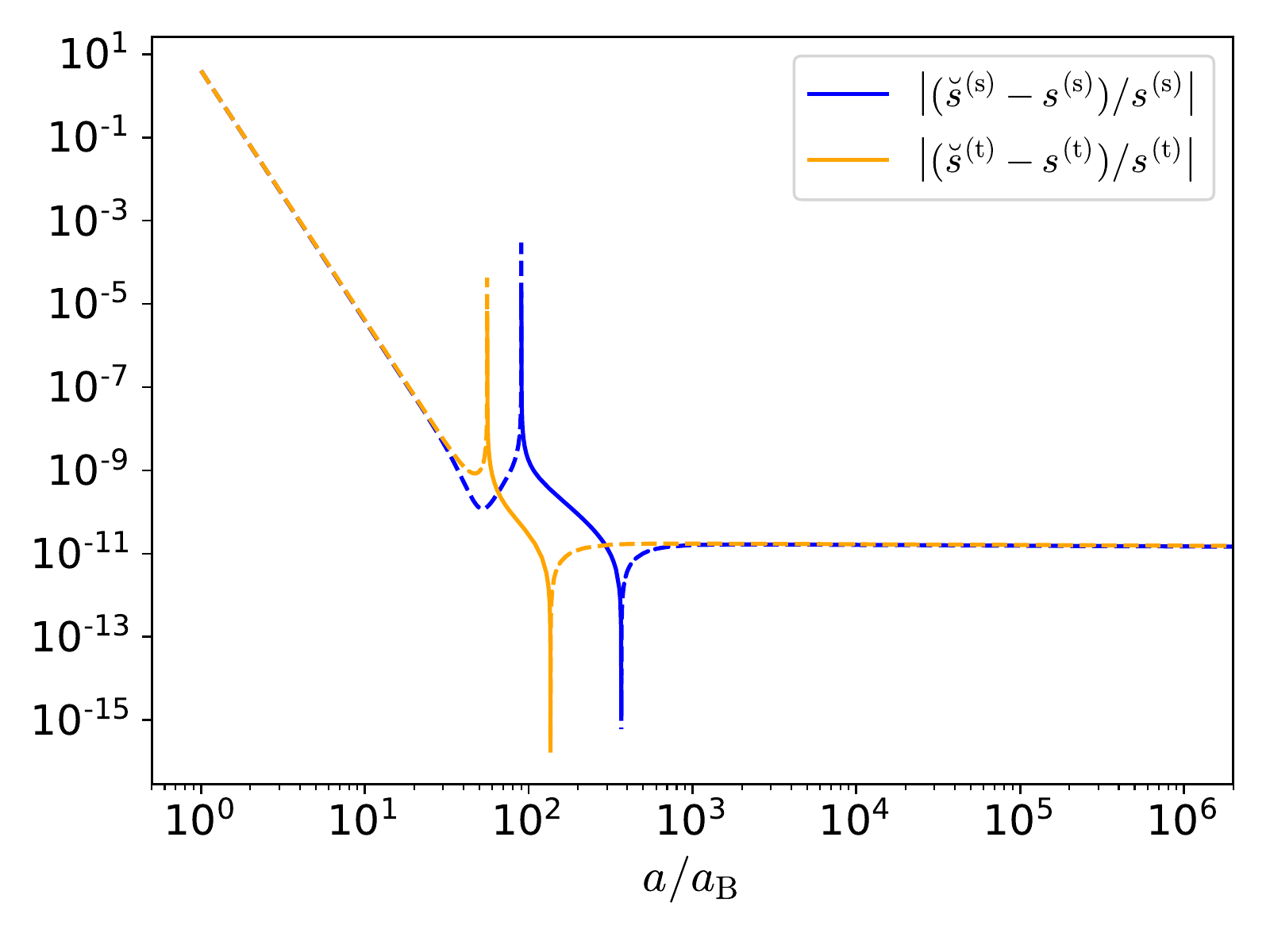} 
	\caption{Left panel: Evolution on the expanding branch of the Universe of the time-dependent masses for the tensor and the Mukhanov-Sasaki perturbations in the hybrid approach, denoted by $s^{({\rm t})}$ and $s^{({\rm s})}$ respectively, and in the dressed metric approach, denoted by $\breve{s}^{({\rm t})}$ and $\breve{s}^{({\rm s})}$, respectively. Right panel: Relative difference between the value of the time-dependent mass in the two approaches, both for the tensor and for the Mukhanov-Sasaki perturbations. In this right panel, solid (dashed) lines correspond to positive (negative) values of the quantities for which we represent the absolute value.
		Here, we consider a quadratic potential, $V(\phi) = m^{2}\phi^2/2$. In Planck units, the inflaton field at the bounce and the parameters of the model are taken equal to $\phi_{\rm B} = 0.97$, $m =1.20\cdot 10^{-6}$, and $\gamma = 0.2375$.}
	\label{fig:tdm}
\end{figure}

In Fig. \ref{fig:uvpot}, we compare the absolute values of the Mukhanov-Sasaki potentials $\mathcal{U}$ and $\mathcal{V}$, as well as their relative values with respect to the corresponding time-dependent mass of the scalar perturbations (in the hybrid and dressed metric approaches, respectively), for one of such kinetically dominated solutions in the case of a quadratic potential for the inflaton. The initial conditions and inflaton mass for this solution were considered in Ref. \cite{GUI}, in the numerical study of the consequences of the hybrid approach for the CMB, where they were shown to lead to power spectra compatible with observations while displaying power suppression and certain superposed features at large scales. In this sense, we recall that, using the initial value $a_{\rm B}$ of the scale factor $a$ as a length scale and imposing the effective homogeneous Hamiltonian, we can reduce the set of initial conditions at the bounce (where the time derivative of the scale factor vanishes) just to the value of the inflaton. In addition, we plot in Fig. \ref{fig:tdm} the time-dependent masses for the tensor and the Mukhanov-Sasaki perturbations, both in the hybrid and the dressed metric cases, to show how tiny the effect of the potentials $\mathcal{U}$ and $\mathcal{V}$ is in this particular kinetically dominated solution. In the same figure, we also plot the relative difference between the value in the two approaches of the time-dependent tensor mass and of the scalar one. For the dressed metric, we have chosen the Mukhanov-Sasaki potential $\mathcal{V}$ with the prescription of sign of Refs. \cite{waco,waco1} for the term proportional to $V_{,\phi}$.

\section{Time-dependent masses at the Big Bounce}\label{sec3}

The equations for the tensor and the Mukhanov-Sasaki perturbations in the hybrid and dressed metric approaches are of the harmonic oscillator type, with different time-dependent masses in each case. In this section we will analyze the properties of these masses when they are evaluated at the big bounce in effective LQC, focusing on their positivity. If one chooses the initial time when the bounce occurs, this positivity may be relevant in the search of well-defined and physically interesting initial conditions for the gauge-invariant perturbations, for instance if one wants to construct sets of initial conditions corresponding to adiabatic states \cite{dressed2,dressed3,dressed4,Morris,hybr-pred,GUI}. 

Let us call
\begin{align}\label{hmass}
s^{({\rm t})}=-\frac{4\pi G}{3}a^2(\rho-3P)\qquad \text{and}\qquad s^{({\rm s})}=s^{({\rm t})}+\mathcal{U}
\end{align}
the time-dependent masses for the tensor and the Mukhanov-Sasaki perturbations in the hybrid approach, respectively. Similarly, we will call
\begin{align}\label{dmass}
\breve{s}^{({\rm t})}=-\frac{4\pi G}{3}a^2\rho\left(1+2\frac{\rho}{\rho_{\rm max}}\right)+4\pi G a^2 P\left(1-2\frac{\rho}{\rho_{\rm max}}\right)\qquad \text{and} \qquad \breve{s}^{({\rm s})}=\breve{s}^{({\rm t})}+\mathcal{V}
\end{align}
the corresponding masses in the dressed metric approach. Recall that $\mathcal{U}$ and $\mathcal{V}$ are given in Eqs. \eqref{Ueff} and \eqref{Veff}, and these homogeneous quantities must be evaluated on effective trajectories.

In effective LQC, the big bounce occurs when the energy density of the inflaton field $\rho$, given in Eq. \eqref{density}, equals its upper bound $\rho_{\rm max}=3/(8\pi G \gamma^{2}\Delta)$. When this happens, the modified Friedmann equation \eqref{LQC} results in a vanishing Hubble parameter and the scale factor reaches its minimum possible value $a_{\rm B}$. In what follows, the symbol B as subscript or superscript of any homogeneous variable stands for its evaluation at the bounce. In this situation, it is easy to check that the time-dependent masses adopt the expressions
\begin{align}\label{hybformul}
\frac{s^{({\rm t})}_{\rm B}}{8\pi G a_{\rm B}^2}=\frac{1}{8\pi G \gamma^2\Delta}- V(\phi_{\rm B})= \frac{\rho_{\rm max}}{3}- V(\phi_{\rm B}), \qquad s^{({\rm s})}_{\rm B}=s^{({\rm t})}_{\rm B}+{\mathcal{U}}_{\rm B},
\end{align}
in the hybrid case, whereas for the dressed metric approach one obtains
\begin{align}
\frac{\breve{s}^{({\rm t})}_{\rm B}}{8\pi G a_{\rm B}^2}=-\frac{3}{8\pi G \gamma^2\Delta}+ V(\phi_{\rm B})= -\rho_{\rm max}+ V(\phi_{\rm B}),\qquad \breve{s}^{({\rm s})}_{\rm B}=\breve{s}^{({\rm t})}_{\rm B}+{\mathcal{V}}_{\rm B}.
\end{align}
In these formulas, the Mukhanov-Sasaki potentials at the bounce are 
\begin{align}
&\mathcal{U}_{\rm B}=a_{\rm B}^2\left[V_{,\phi\phi}^{\rm B}+48\pi G V(\phi_{\rm B})-128\pi^2 G^2 \gamma^2 \Delta V^{2}(\phi_{\rm B})\right],\\&\label{Vbd}\mathcal{V}_{\rm B}=a_{\rm B}^2\left[V_{,\phi\phi}^{\rm B}+48\pi G V(\phi_{\rm B})-\text{sign}\left(\sqrt{\mathfrak{f}}\right)16\pi G \gamma \sqrt{\Delta}\frac{|{\phi}'_{\rm B}|}{a_{\rm B}}V_{,\phi}^{\rm B}-128\pi^2 G^2 \gamma^2 \Delta V^{2}(\phi_{\rm B})\right].
\end{align}
The most relevant differences between the time-dependent masses for the two quantization approaches, commented in the previous section, show up here. In particular, these differences are important when one analyzes the positivity of these masses at the bounce. 

As we have discussed in the Introduction, in the regime of very high kinetic dominance, where the effects of the potential can be safely ignored, one can use e.g. the analytic solution obtained in Ref. \cite{waco} to elucidate the sign of the masses at the bounce for each of the two considered approaches. Nevertheless, as we have explained, our goal is to exceed this regime and carry out a more general analysis in which the influence of the potential in the positivity of the mass can be quantified. Of course, the conclusions that we will reach in this way will, in particular, reproduce in the limit of vanishing potential the results for solutions with an inflaton energy density that is totally dominated by its kinetic contribution.

For the tensor perturbations, it is straightforward to deduce from Eq. \eqref{hybformul} that, in the hybrid approach, the mass at the bounce, $s^{({\rm t})}_{\rm B}$, is positive if and only if $V(\phi_{\rm B})< \rho_{\rm max}/3$. Notice that this upper bound on the potential is compatible with kinetic dominance at the bounce, since it suffices that the kinetic contribution to the energy density of the inflaton is larger than $2\rho_{\rm max}/3$, and hence always larger than twice the potential. Obviously, this kinetic contribution is bounded from above by $\rho_{\rm max}$ for non-negative potentials. In the case of the dressed metric approach, on the other hand, the mass at the bounce for the tensor perturbations is never positive, because the potential is necessarily smaller than, or equal to, the upper bound for the energy density. This nonpositivity was certainly expected, because the analyzed tensor mass is known to coincide with the effective value of $-a''/a$ in the dressed metric approach. Since the dressed scale factor has a minimum at the bounce, its second derivative is non-negative and then, trivially, the studied ratio cannot be positive.

The analysis of the Mukhanov-Sasaki masses is more involved, because it depends on the explicit form of the inflaton potential $V(\phi)$. In order to perform this analysis, we will treat the value of $V_{,\phi\phi}^{\rm B}$ as a parameter, that we will assume non-negative. This is what happens in fact for the massive scalar field, for which the second derivative of the potential is just a positive constant, namely the square mass of the inflaton. Besides, we will restrict our attention to potentials that are non-negative, like the one for the massive scalar field. Therefore, at the bounce we must have $0\leq V(\phi_{\rm B})\leq \rho_{\rm max}$. 

In the hybrid approach, the time-dependent mass for the Mukhanov-Sasaki perturbations takes the form of a quadratic polynomial in $V(\phi_{\rm B})$. The roots of this polynomial are
\begin{align}
x_{\pm}=\frac{5\pm\sqrt{33+8\gamma^2\Delta V_{,\phi\phi}^{\rm B}}}{12}\rho_{\rm max},
\end{align}
and the polynomial decreases for large values of $V(\phi_{\rm B})$. Given our assumption of a non-negative second derivative of the inflaton potential, $x_{-}$ is clearly negative, and $x_+$ positive. Consequently, for non-negative potentials, the mass $s^{({\rm s)}}_{\rm B}$ is   positive if and only if $V(\phi_{\rm B})< x_+$. The availability of a sector of kinetic dominance at the bounce with positive time-dependent mass is then directly granted, since this positivity holds for sufficiently small potentials and the kinetic contribution to the energy density at the bounce is simply $\rho_{\rm max}-V(\phi_{\rm B})$. Moreover, since $x_+$ is always larger than its value for vanishing $V_{,\phi\phi}^{\rm B}$, which equals $(5+\sqrt{33})\rho_{\rm max}/12$, the interval $[x_+,\rho_{\rm max}]$ of potentials at the bounce below the upper bound for the energy density is either empty or included in $[(5+\sqrt{33})\rho_{\rm max}/12,\rho_{\rm max}]$. Therefore, since $(5+\sqrt{33})/12\approx 0.895$, it is only at most in a relatively restricted interval of large potentials away from the kinetically dominated sector that the mass might not be positive in the effective regime of hybrid LQC.
	
In the dressed metric approach, the Mukhanov-Sasaki mass at the bounce, $\breve{s}^{({\rm s})}_{\rm B}$, contains the new factor $V_{,\phi}^{\rm B}$, something that adds an extra complication to the analysis of the positivity. In order to complete the study analytically, apart from the already assumed non-negativity of the inflaton potential and of its second derivative at the bounce, we will suppose that the first derivative at the bounce can be bounded in the form $|V_{,\phi}^{\rm B}|\leq C \sqrt{2 V^{\rm B}_{,\phi\phi} V(\phi_{\rm B})}$, where $C\equiv C(V^{\rm B}_{,\phi\phi})$ may be any positive bounded function of the order of the unit. This assumption of a bound is not too restrictive, and in particular contains the relevant case of the quadratic potential, that satisfies the functional relation $|V_{,\phi}|=\sqrt{2V_{,\phi\phi}V(\phi)}$ for all values of the inflaton at all instants of time, and not only at the bounce. Thus, in this case, one can make $C=1$.  On the other hand, given the relation of the energy density with the time derivative of the inflaton and its potential [see Eq. \eqref{density}], evaluated at the bounce where $\rho=\rho_{\rm max}$, one obtains that $|\phi'_{\rm B}| \sqrt{2V(\phi_{\rm B})}/a_{\rm B}\leq \rho_{\rm max}$. With all this information, it is easy to deduce that
\begin{align}
16\pi G \gamma\sqrt{\Delta}\frac{|\phi'_{\rm B} V^{\rm B}_{,\phi}|}{a_{\rm B}}\leq\frac{6 C}{\gamma\sqrt{\Delta}}\sqrt{V_{,\phi\phi}^{\rm B}}.
\end{align}
Noting that  $V^{\rm B}_{,\phi}$, and hence the term proportional to it in $\mathcal{V}_{\rm B}$, may take both signs, one concludes that the Mukhanov-Sasaki mass at the bounce in the dressed metric approach is bounded from below and from above by $P_{-}\leq \breve{s}^{({\rm s})}_{\rm B} \leq P_{+}$, where $P_{-}$ and $P_{+}$ are the following quadratic polynomials in $V(\phi_{\rm B})$:
\begin{align}\label{u}
P_{\pm}=\breve{s}_{\rm B}^{({\rm t})}+a_{\rm B}^2\left[V_{,\phi\phi}^{\rm B}+48\pi G V(\phi_{\rm B})\pm\frac{6C}{\gamma\sqrt{\Delta}}\sqrt{V_{,\phi\phi}^{\rm B}}-128\pi^2 G^2 \gamma^2 \Delta V^{2}(\phi_{\rm B})\right].
\end{align}
Both polynomials $P_{\pm}$ decrease for large $|V(\phi_{\rm B})|$. Their roots, $y_{\pm}(P_{+})$ and $y_{\pm}(P_{-})$, respectively, are given by
\begin{align}\label{roots}	y_{\pm}(P_{\pm})=\frac{7\pm\sqrt{25+8\gamma^2\Delta V_{,\phi\phi}^{\rm B}\pm 48 \gamma C (\Delta V^{\rm B}_{,\phi\phi})^{1/2}}}{12}\rho_{\rm max}.
\end{align}
For non-negative $V^{\rm B}_{,\phi\phi}$, the two roots of $P_+$ are always real. Hence, the bound $\breve{s}^{({\rm s})}_{\rm B} \leq P_{+} $ implies that the mass is negative outside of the interval $[y_{-}(P_{+}),y_{+}(P_{+}) ]$ for $V(\phi_{\rm B})$ in the sector of physical interest $[0,\rho_{\rm max}]$. It is clear that $y_{-}(P_{+})< \rho_{\rm max}$. Then, the Mukhanov-Sasaki mass of the dressed metric approach is not positive in the subinterval $[0,y_{-}(P_{+})]$ provided that $y_{-}(P_{+})>0$. This last condition is satisfied for $\gamma^{2}\Delta V_{,\phi\phi}^{\rm B}<(\sqrt{9 C^2+3} -3 C)^2$. In particular, this includes the sector of small $\gamma^{2}\Delta V^{\rm B}_{,\phi\phi}$. It is worth commenting that, for the quadratic potential and with the typical values of the inflaton mass that are favored phenomenologically in order to get power spectra compatible with the observations of the CMB and still presenting power suppression at large scales \cite{GUI}, one has that $\gamma^2 \Delta V^{\rm B}_{,\phi\phi}$ is as small as $10^{-12}$. Let us also point out that the interval $[0,y_{-}(P_{+})]$ in which the mass becomes nonpositive grows as large as $[0,\rho_{\rm max}/6]$ in the limit in which $V_{,\phi\phi}^{\rm B}$ tends to zero. So, in those cases where the root $y_{-}(P_+)$ is positive, something that occurs if $V_{,\phi\phi}^{\rm B}$ is not very large and certainly in the physically interesting region of very small values of $V_{,\phi\phi}^{\rm B}$, the studied mass is inevitably negative for $V(\phi_{\rm B})$ in a nonempty neighborhood of zero, which is precisely the region containing the solutions that are kinetically dominated.

Finally, if not only the roots of $P_+$, but also those of $P_-$ are real, one can straightforwardly check that
\begin{equation} 
  y_{-}(P_{+})\leq y_{-}(P_{-}) \leq y_{+}(P_-)\leq y_{+}(P_+). 
\end{equation}
This reality of all roots occurs if $\gamma^2 \Delta V^{\rm B}_{,\phi\phi}\geq (3C +\sqrt{9 C^2-25/8}\, )^2 $ or if $\gamma^2 \Delta V^{\rm B}_{,\phi\phi}\leq (3C -\sqrt{9 C^2-25/8}\, )^2 $, which includes the region of small values of $V^{\rm B}_{,\phi\phi}$. Taking then into account the bound $P_-\leq \breve{s}^{(s)}_B $, we can be sure that the mass is non-negative at least in the intersection of $[y_{-}(P_{-}),y_{+}(P_{-}) ]$ with the interval $[0,\rho_{\rm max}]$, to which $V(\phi_{\rm B})$ is restricted. 

\begin{figure}
 \includegraphics[width=0.49\textwidth]{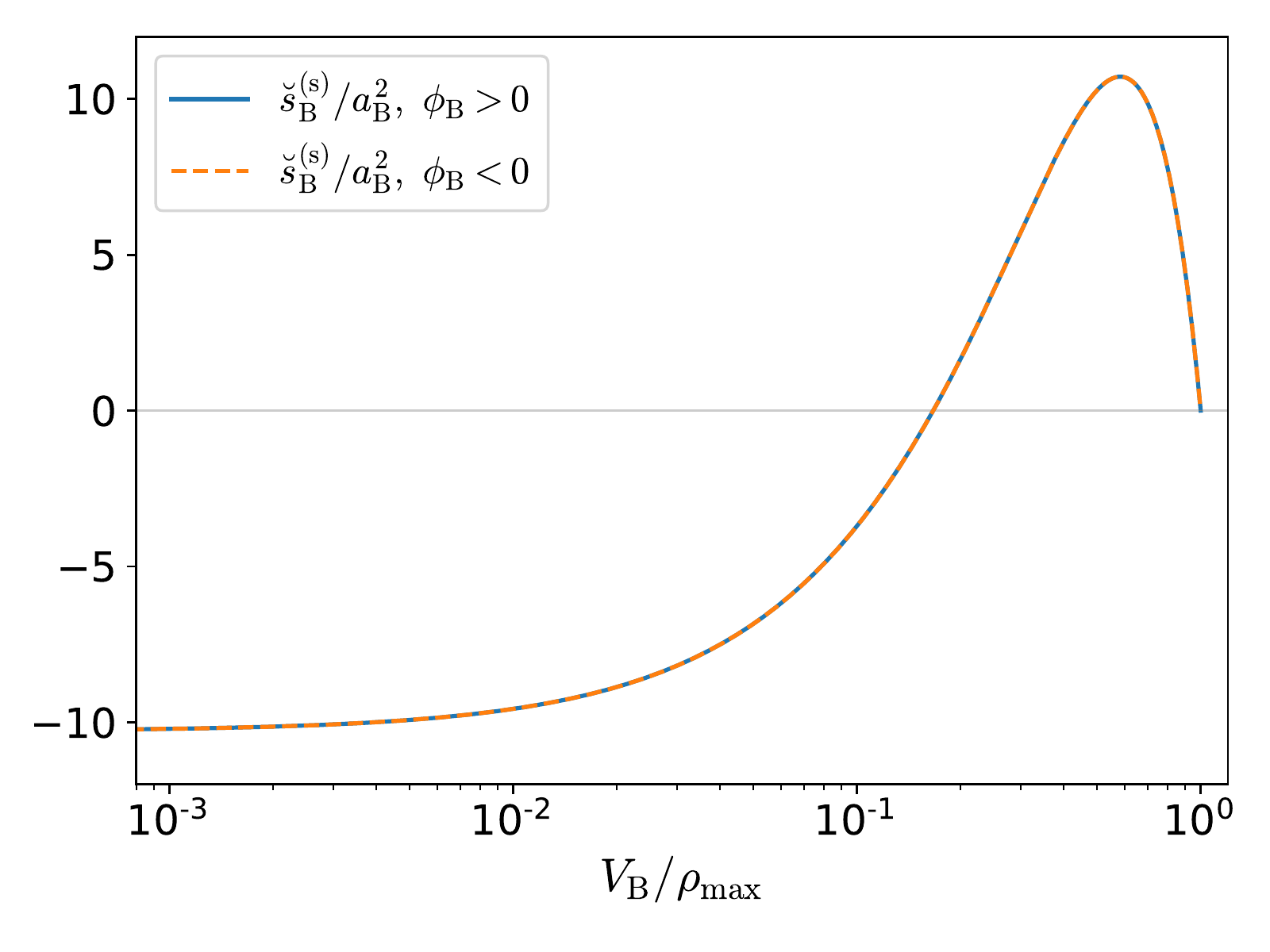}
 \includegraphics[width=0.49\textwidth]{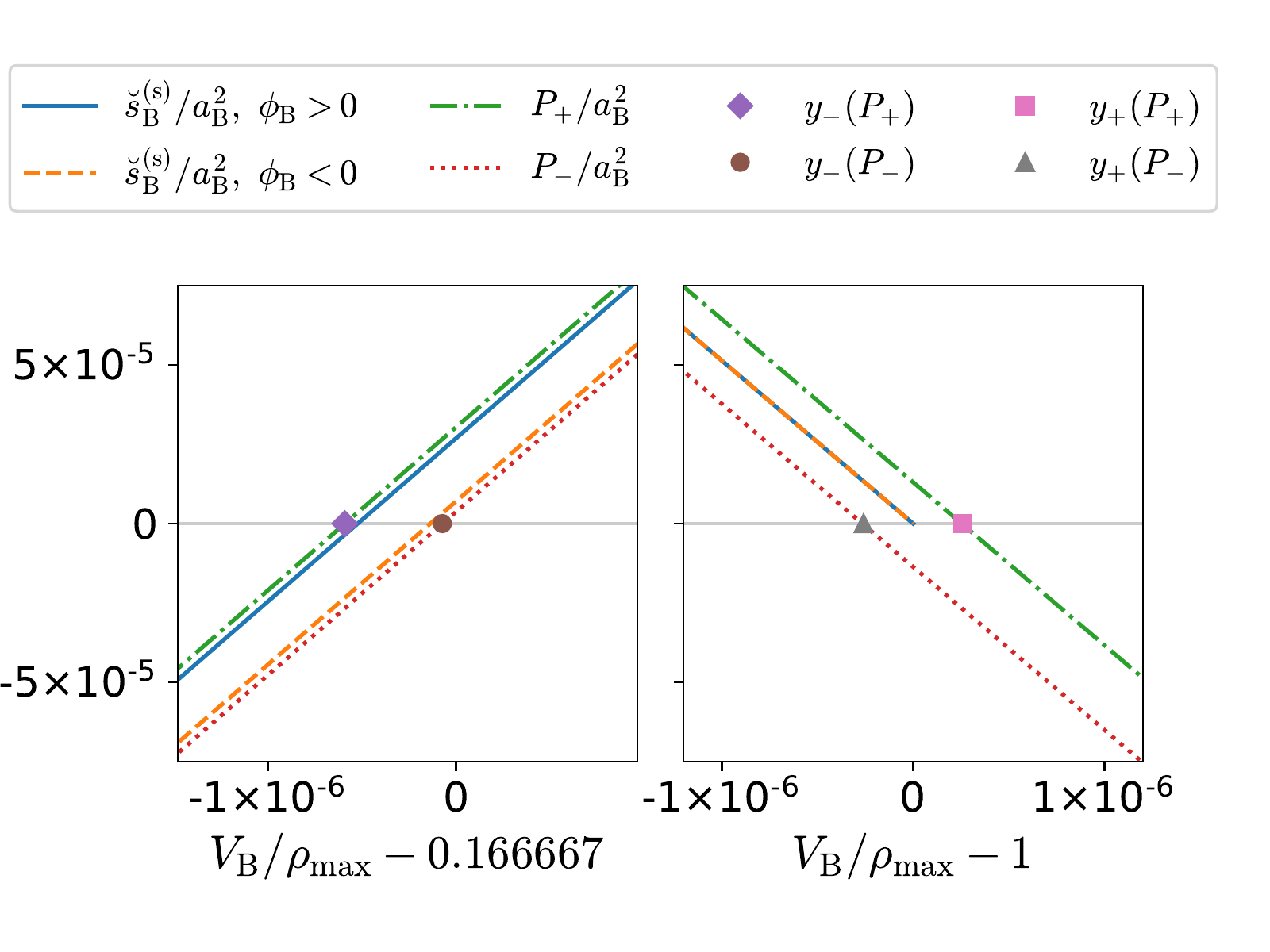}
 \caption{Left panel: The time-dependent Mukhanov-Sasaki mass at the bounce in the dressed metric approach as a function of the value of the potential at the bounce, for the case of the quadratic potential, $V(\phi) = m^{2}\phi^2/2$, with $m =1.20\cdot10^{-6}$ in Planck units. Again, we take $\gamma=0.2375$. We consider the two possibilities of positive and negative values of the inflaton at the bounce.  Right panel: Zoom of the two regions in which the time-dependent mass changes sign. In these regions, we also plot the polynomials $P_{\pm}$ and their zeros.}
 \label{fig:msdm}
\end{figure}

In Fig. \ref{fig:msdm} we plot the value of the time-dependent mass for the Mukhanov-Sasaki perturbations in the dressed metric approach particularized to the case of the quadratic potential, with the same value of the inflaton mass as in Figs. \ref{fig:uvpot} and \ref{fig:tdm}. We zoom the regions where the mass changes from positive to negative values, comparing those points with the roots of the polynomials $P_{\pm}$. Again, we have employed the prescription adopted in Refs. \cite{waco,waco1} for the sign of $\sqrt{\mathfrak{f}}$. 

\section{Conclusions}\label{concl}

We have considered the field equations for the gauge-invariant scalar and tensor perturbations arising from two approaches to LQC: the hybrid and the dressed metric approaches. In order to apply the loop quantization of the geometry, they both combine a polymeric quantization of the background with a Fock quantization of the cosmological perturbations. Moreover, in both cases, one obtains field equations with a standard hyperbolic behavior in the ultraviolet regions, the quantum effects on the geometry being incorporated as a modification of the time-dependent mass of the perturbations with respect to GR. We have focused our discussion on the scenarios that have received more attention in the literature, namely, the situations without backreaction of the perturbations and with background geometries that are describable in terms of effective LQC. We have seen that, in spite of the great similarities between the two approaches, the time-dependent masses deduced with them differ. In the studied scenarios, this is the only noticeable difference, and can be understood as a consequence of the distinct quantization procedures that are followed in each of the approaches. 

The hybrid approach treats the whole system composed by the background and the perturbations as a symplectic constrained system, after truncating its action at quadratic order in the perturbations. The formulation is canonical, and the effective description of LQC is incorporated only at the end, once all background quantities are expressed in terms of the basic variables. The dressed metric approach, on the other hand, deals first with the background, and lifts its effective trajectories to the truncated phase space that contains the perturbations, incorporating the effects of the quantization of the geometry precisely by replacing the classical metric with a dressed one. In this framework, it is the time derivatives of the dressed metric that are incorporated in the equations for the perturbations. Since these derivatives are computed within the effective dynamics, their relation with the canonical variables of the geometry departs from the standard one in GR. This is the main reason behind the difference between the time-dependent masses of the two approaches. The part of those masses that is common for the tensor and scalar perturbations equals $- a^{\prime\prime}/a$ in GR. The two distinct procedures by which one quantizes the second derivative of the background scale factor lead to the noticed discrepancy between the considered masses. 

There is a related but more subtle difference between the time-dependent masses in the two studied cases, precisely in the additional term that appears for the Mukhanov-Sasaki perturbations, that contains the dependence on the inflaton potential and its two first derivatives. More specifically, this difference affects the contribution that is proportional to the first derivative of the potential. The distinction is due to the quantization prescription adopted for a factor of the form $1/b$, where $b$ is a classical variable proportional to the homogeneous Hubble parameter (see, e.g., Ref. \cite{hybr-inf5}). In the hybrid quantization, the requirement of preserving the superselection sectors of the background geometry at the moment of defining the action of the Hamiltonian constraint of the entire system (i.e., the background plus the perturbations), leads to an effective counterpart of the type $\sin{(2b)}/(2 \sin^2{b})$. In the dressed metric approach, on the other hand, the truncated phase space is not constrained as a whole, and this factor is made to correspond to the square root of $1/\sin^2{b}$ in the effective description. Nonetheless, given that the term where this discrepancy appears is proportional to the derivative of the inflaton potential, this additional difference turns out to be generically negligible for solutions where the energy density of the inflaton is clearly dominated by the kinetic contribution, at least compared to the other differing part of the time-dependent masses that we have found.

We have studied in detail the properties of the masses of the two approaches at the big bounce experimented by the effective background. This bounce marks a special instant of time, when the Hubble constant vanishes. It seems reasonable to consider that instant as a natural choice of initial time, in which one can fix initial conditions for the perturbations. In the definition of those initial conditions, the properties of the time-dependent mass, and in particular its positivity, can be very important, for instance, if one wants to determine data that correspond to adiabatic states \cite{dressed2,dressed3,dressed4,Morris,hybr-pred,GUI} away from the ultraviolet region. We have seen, however, that the mass of the tensor perturbations is never positive in the dressed metric approach. For the hybrid approach, on the contrary, we have proven the positivity of the tensor mass in the sector of background solutions for which the inflaton energy density is dominated by the kinetic term. Furthermore, non-negativity is granted if the kinetic contribution lies in the interval $[2 \rho_{{\rm max}}/3, \rho_{{\rm max}} ]$, where $\rho_{{\rm max}}$ is the upper bound on the inflaton energy density, saturated at the bounce. 

The analysis at the bounce of the positivity of the time-dependent mass for the Mukhanov-Sasaki gauge-invariant perturbations is more involved, owing to the appearance of the inflaton potential and its derivatives in the corresponding expression. Restricting the study to non-negative potentials with a non-negative second derivative at the bounce, and treating this second derivative as a parameter, we have proven that the mass is not negative in the hybrid approach at least for all background solutions with a kinetic energy of the inflaton at the bounce in $[ (7-\sqrt{33})\rho_{\rm max}/12,\rho_{\rm max}]$, an interval which clearly contains the region of kinetic dominance. In particular, the result is valid for the quadratic potential $V(\phi)=m^2\phi^2/2$, which is non-negative and has a positive second derivative equal to the constant $m^2$, i.e. the squared mass of the inflaton. For the dressed metric, on the other hand, the time-dependent gauge-invariant scalar mass includes an additional term that is proportional to the first derivative of the inflaton potential. To deal with it without introducing unnecessary complications, we have further restricted the discussion to potentials that satisfy the relation $|V^{\rm B}_{,\phi}|\leq C \sqrt{2 V^{\rm B}_{,\phi\phi}V(\phi_{\rm B})}$, where $C$ can be any positive bounded function of $V^{\rm B}_{,\phi\phi}$ of the order of the unit. The analysis includes again the interesting case of the quadratic potential, for which this relation holds as an equality with $C=1$. We have then demonstrated that, if $\gamma^{2}\Delta V_{,\phi\phi}^{\rm B}<(\sqrt{9 C^2+3} -3 C)^2$, where $\gamma$ is the Immirzi parameter and $\Delta$ the area gap allowed by LQG, there exists an interval of kinetic energies for the inflaton at the bounce for which the Mukhanov-Sasaki mass is negative. This interval always contains a neighborhood of $\rho_{\rm max}$, and therefore includes the sector of kinetically dominated solutions. For the case of the quadratic potential and values of the inflaton mass favored phenomenologically in LQC, in order to derive power spectra compatible with observations that nonetheless contain traces of quantum effects at large scales, the value of  $\gamma^{2}\Delta V_{,\phi\phi}^{\rm B}$ is really small, around $10^{-12}$ or less, so that the above condition is clearly satisfied. Furthermore, for such almost negligible values of the second derivative of the potential at the bounce, the interval of kinetic energies for which the mass is negative is very approximately equal to $[5 \rho_{\rm max}/6, \rho_{\rm max}]$. 

\acknowledgments

The authors are grateful to J. Olmedo for very enlightening discussions. This work was supported by Project. No. MINECO FIS2014-54800-C2-2-P from Spain and its continuation Project. No. MINECO FIS2017-86497-C2-2-P.

\end{document}